\documentclass[journal=jacsat,manuscript=article]{achemso}
\usepackage[version=3]{mhchem} 

\title{Radio-frequency reflectometry in bilayer graphene devices utilizing micro graphite back-gates}

\author{Tomoya Johmen}
\affiliation{Research Institute of Electrical Communication, Tohoku University, 2-1-1 Katahira, Aoba-ku, Sendai 980-8577, Japan}
\affiliation{Department of Electronic Engineering, Tohoku University, Aoba 6-6-05, Aramaki, Aoba-Ku, Sendai 980-8579, Japan}

\author{Motoya Shinozaki}
\affiliation{Research Institute of Electrical Communication, Tohoku University, 2-1-1 Katahira, Aoba-ku, Sendai 980-8577, Japan}
\affiliation{Department of Electronic Engineering, Tohoku University, Aoba 6-6-05, Aramaki, Aoba-Ku, Sendai 980-8579, Japan}

\author{Yoshihiro Fujiwara}
\affiliation{Research Institute of Electrical Communication, Tohoku University, 2-1-1 Katahira, Aoba-ku, Sendai 980-8577, Japan}
\affiliation{Department of Electronic Engineering, Tohoku University, Aoba 6-6-05, Aramaki, Aoba-Ku, Sendai 980-8579, Japan}

\author{Takumi Aizawa}
\affiliation{Research Institute of Electrical Communication, Tohoku University, 2-1-1 Katahira, Aoba-ku, Sendai 980-8577, Japan}
\affiliation{Department of Electronic Engineering, Tohoku University, Aoba 6-6-05, Aramaki, Aoba-Ku, Sendai 980-8579, Japan}

\author{Tomohiro Otsuka}
\email{tomohiro.otsuka@tohoku.ac.jp}
\affiliation{WPI Advanced Institute for Materials Research, Tohoku University, Sendai 980-8577, Japan}
\affiliation{Research Institute of Electrical Communication, Tohoku University, 2-1-1 Katahira, Aoba-ku, Sendai 980-8577, Japan}
\affiliation{Department of Electronic Engineering, Tohoku University, Aoba 6-6-05, Aramaki, Aoba-Ku, Sendai 980-8579, Japan}
\alsoaffiliation{Center for Spintronics Research Network, Tohoku University, 2-1-1 Katahira, Aoba-ku, Sendai 980-8577, Japan}
\alsoaffiliation{Center for Science and Innovation in Spintronics, Tohoku University, 2-1-1 Katahira, Aoba-ku, Sendai 980-8577, Japan}
\alsoaffiliation{Center for Emergent Matter Science, RIKEN, 2-1 Hirosawa, Wako, Saitama 351-0198, Japan}


\begin{document}
\begin{abstract}
Bilayer graphene is an attractive material that realizes high-quality two-dimensional electron gas with a controllable bandgap.
By utilizing the bandgap, electrical gate tuning of the carrier is possible and formation of nanostructures such as quantum dots have been reported.
To probe the dynamics of the electronics states and realize applications for quantum bit devices, RF-reflectometry which enables high-speed electric measurements is important.
Here we demonstrate RF-reflectometry in bilayer graphene devices.
We utilize a micro graphite back-gate and an undoped Si substrate to reduce the parasitic capacitance which degrades the RF-reflectometry.
We measure the resonance properties of a tank circuit which contains the bilayer graphene device.
We form RF-reflectmetory setup and compared the result with the DC measurement, and confirmed their consistency.
We also measure Coulomb diamonds of quantum dots possibly formed by bubbles and confirm that RF-reflectometry of quantum dots can be performed.
This technique enables high-speed measurements of bilayer graphene quantum dots and contributes to the research of bilayer graphene-based quantum devices by fast readout of the states.
\end{abstract}

Bilayer graphene (BLG) is a nanocarbon material that has attracted much attention as a new material with unique properties~\cite{zhang2009direct,dean2010boron}.
With this new material, formation of quantum dots and exploring those properties have been intensively studied~\cite{2009MoriyamaNanoLett, 2011GutingerPRB, 2011VolkNanoLett, 2018EichPRX, kurzmann2019charge,kurzmann2019excited, banszerus2020single, banszerus2021pulsed, banszerus2021spin}.
Quantum dots~\cite{tarucha1996shell, kouwenhoven1997excitation, kouwenhoven2001few, 2007HansonRMP} are nowadays essential elements in semiconductor spin quantum bits~\cite{1998LossPRA, 2005PettaSci, 2006KoppensNature, 2010LaddNature, 2015VeldhorstNat, 2016OtsukaSciRep, 2018YonedaNatNano} and electronic sensors~\cite{2010AltimirasPRL, 2015OtsukaSciRep, 2019OtsukaPRB}. 
BLG has the potential to improve the performance by utilizing the weak spin-orbit interaction and decreasing the nuclear spins~\cite{2007TrauzettelNatPhys}.

In BLG, most of the reports directly measure the current flowing through the quantum dots with low-frequency electronics.
To access the faster dynamics of the electronic states and for applications to quantum bits, faster and more sensitive electronic measurements are required.
In order to realize this faster readout, radio-frequency (RF) reflectometry is a powerful tool to improve the measurement bandwidth~\cite{schoelkopf1998radio, qin2006radio, reilly2007fast, barthel2009rapid, 2010BarthelPRB}.
They apply RF signals to resonators including the target devices and monitor the reflected RF signals.
The impedance of the target devices can be monitored through the reflected signals.
In this measurement setup, we need to decrease the stray capacitance in the measurement circuit to reduce the RF leakage and optimize the properties of the resonators.
But in the conventional BLG devices on SiO$_{2}$ and highly doped Si back-gates, big stray capacitances are formed by the device electrodes and the back-gates.

Various techniques have been developed to reduce the stray capacitance in graphene devices~\cite{banszerus2021dispersive}, such as electrolyte gates~\cite{fu2014electrolyte} and matching network circuits~\cite{muller2012fast}.
Among them, reducing the size of the capacitance formed by the gate electrode is a simple approach.
Such an approach is reported in Si-based quantum dots~\cite{noiri2020radio, mizokuchi2021radio, 2021BuguJJAP}.
In this Letter, we realize RF-reflectometry in BLG devices utilizing micro graphite back-gates to reduce the stray capacitance.
We demonstrate the operation of the reflectometry and analyze the readout noise.
Furthermore, Coulomb diamonds are observed through the RF-reflectometry.
We consider that quantum dots are formed in the conduction channel by potential fluctuations by bubbles~\cite{bahamon2015conductance, leconte2017graphene}.
These are important in exploring the quantum dynamics in BLG and applications like quantum information processing and readout of the floating gate-type nonvolatile flash memory~\cite{bertolazzi2019nonvolatile}.


\begin{figure}
\begin{center}
  \includegraphics{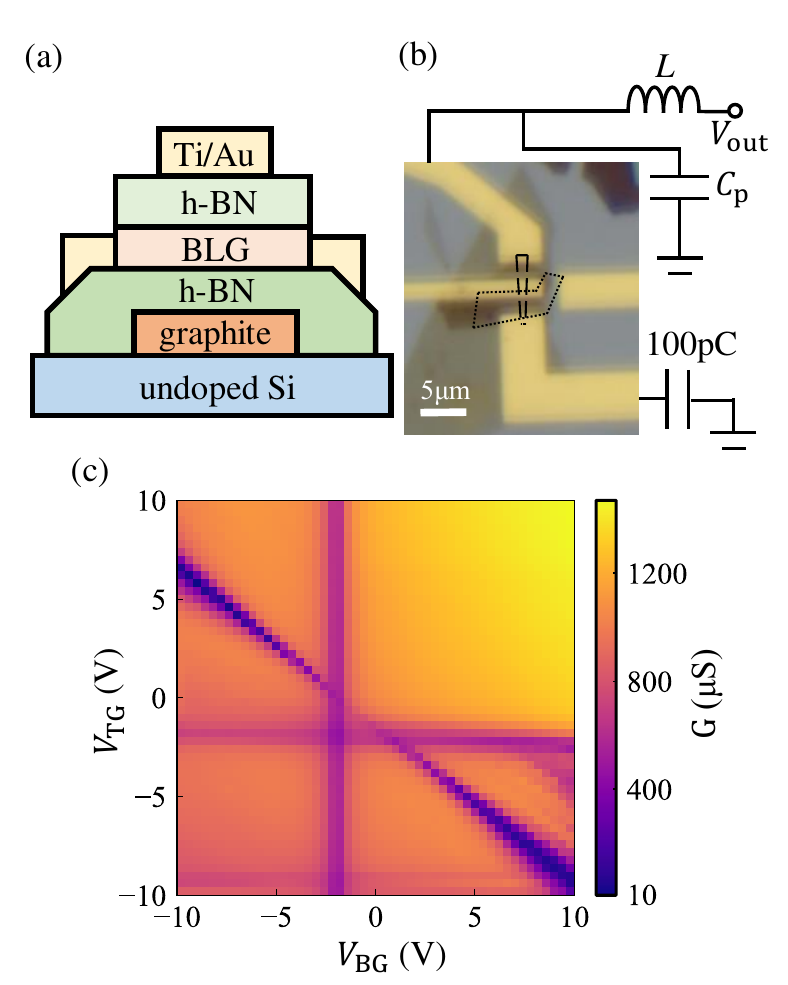}
  \caption{
  (a) Layer structure of the device. The 2DEG is formed in the bilayer graphene which is connected to Ti/Au. The thickness of hexagonal boron nitrides is 30nm. (b) Optical microscope image of the device and schematic of the measurement setup. An RF tank circuit is connected to the source of the device. (c) Observed condactance of the bilayer graphene as a function of $V_{\rm TG}$ and $V_{\rm BG}$.}
  \label{IVsetup}
\end{center}
\end{figure}

Figure~\ref{IVsetup} shows the fabricated device structure.
The device consists of layered 2D materials: a graphite back-gate, a hexagonal boron nitride (hBN), BLG flakes, and another hexagonal boron nitride (hBN) from bottom to top as shown in Fig.~\ref{IVsetup}(a). 
This layered material is placed on top of the undoped Si substrate.
The structure is prepared by the Elvasite transfer method~\cite{2018MasubuchiNatCom}.
On top of the structure, a 2~${\mu}$m wide Ti/Au top gate is placed on hexagonal boron nitride and Ti/Au source and drain contacts are prepared on BLG.
Figure~\ref{IVsetup}(b) shows a photograph of the device taken with an optical microscope. The long dotted line indicates BLG, and the short dotted line indicates graphite back-gate.
Important points of this device are the use of undoped Si substrates and micro graphite back-gates to make the stray capacitance as small as possible. When a sample is placed on a conventional SiO$_{2}$/doped-Si substrate, the capacitance between the contact electrodes and the back-gate becomes large, and RF signal used for reflectometry leaks to the back-gate. By using an undoped Si substrate, the parasitic capacitance is reduced and we are able to create a sample that can be used in RF-reflectometry measurement.

Figure~\ref{IVsetup}(c) shows the conductance of the BLG when the dual gate voltages on the top gate ($V_{\rm TG}$) and the bottom gate ($V_{\rm BG}$) are applied. In this measurement, the device was probed by DC current using a source measurement unit. 
The measurement temperature was 4.2~K.
A low conductance region appears in the diagonal direction.
The total gate voltage $V_{\rm TG}$+$V_{\rm BG}$ controls the Fermi level, and setting the level at near the bandgap leads to the low conductance.
The slope of the diagonal low conductance region is not 1 because there are differences in the electrostatic coupling between $V_{\rm TG}$ and BLG, and $V_{\rm BG}$ and BLG.
Also, the width of the low conductance region reflects the opening of the bandgap in BLG.
The gap is opened by an electric field $\propto |V_{\rm TG}$-$V_{\rm BG}|$~\cite{zhang2009direct}, resulting in the width being very narrow near the origin.
The opening of the bandgap and the accompanying pinch-off properties are observed.
Note that other conductance decreases are observed in holizontal and vertical directions in the graph.
These are the results of charge neutral points by the top and back gate electrodes.
Due to the device structure, there is a region where only the top or back gate covers and this induces the additional decreases in the conductance.


\begin{figure}
\begin{center}
  \includegraphics{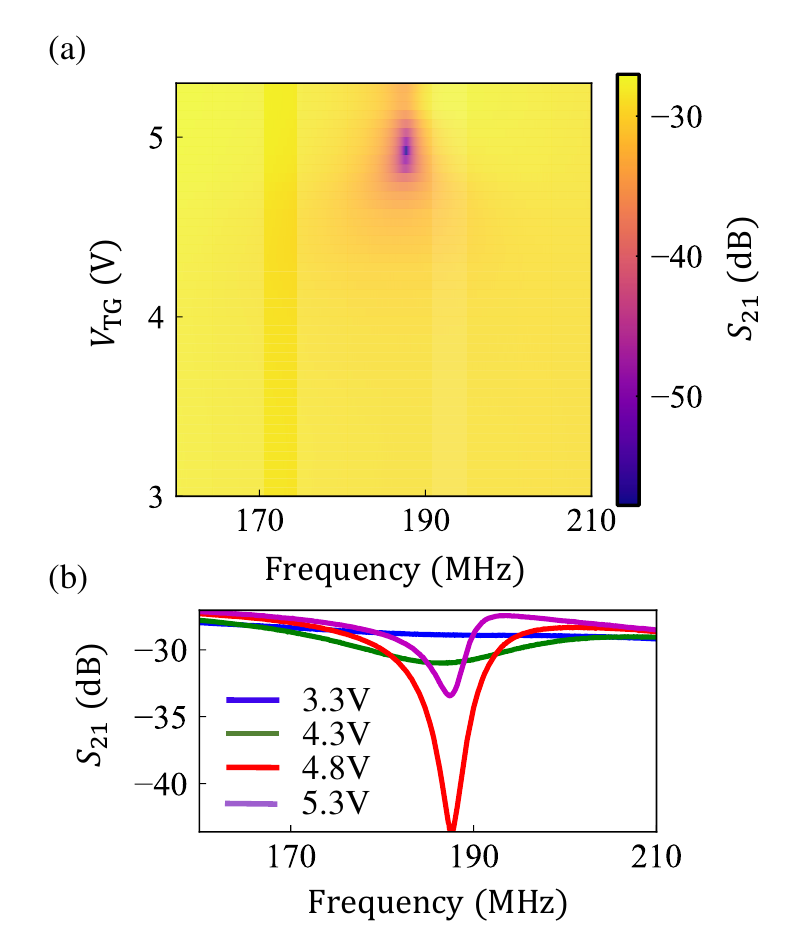}
  \caption{
  (a) Observed $S_{21}$ as a function of the frequency and $V_{\rm TG}$.
  $V_{\rm BG}$ is fixed to -8~V in this measurement.
  The resonance changes with $V_{\rm TG}$.
  (b) Resonance properties at $V_{\rm TG}$ = 3.3, 4.3, 4.8, and 5.3~V.
  }
  \label{VNA}
\end{center}
\end{figure}

Next, we measure the resonance property of a resonator which includes the BLG device.
The RF tank circuit is formed by a chip inductor $L$ = 1.2~$\mu $H, a parasitic capacitor in the circuit $C_{\rm p}$ and the graphene device as shown in Fig~\ref{IVsetup} (b).
We measure the reflected signal from the resonator by a network analyzer through a directional coupler and an amplifier placed in 4.2~K.
Figure~\ref{VNA} (a) shows the observed $S_{21}$ as a function of the frequency and $V_{\rm TG}$.
In this measurement, $V_{\rm BG}$ is fixed to -8~V.
We observe the resonance around $f_{\rm res}$ = 187.6~MHz and the resonance is modified by the change of the device conductance through $V_{\rm TG}$.

Figure~\ref{VNA} (b) shows the resonance traces when $V_{\rm TG}$ = 3.3, 4.3, 4.8, and 5.3~V.
The change of the resonance around $V_{\rm TG}$ = 5~V is consistent with the pinch-off property observed in Fig~\ref{IVsetup} (c). 
The reflected RF signal is modified and RF-reflectometry is working.
The parasitic capacitance is evaluated as $C_{\rm p}$ = 0.60~pF which mainly comes from the parasitic capacitance formed in the measurement board.
The capacitance expected in the BLG device is calculated as 5.4~fF from the geometry.
This value is small compared to 0.60~pF and will not mainly affect the observed resonance.


\begin{figure}
\begin{center}
  \includegraphics{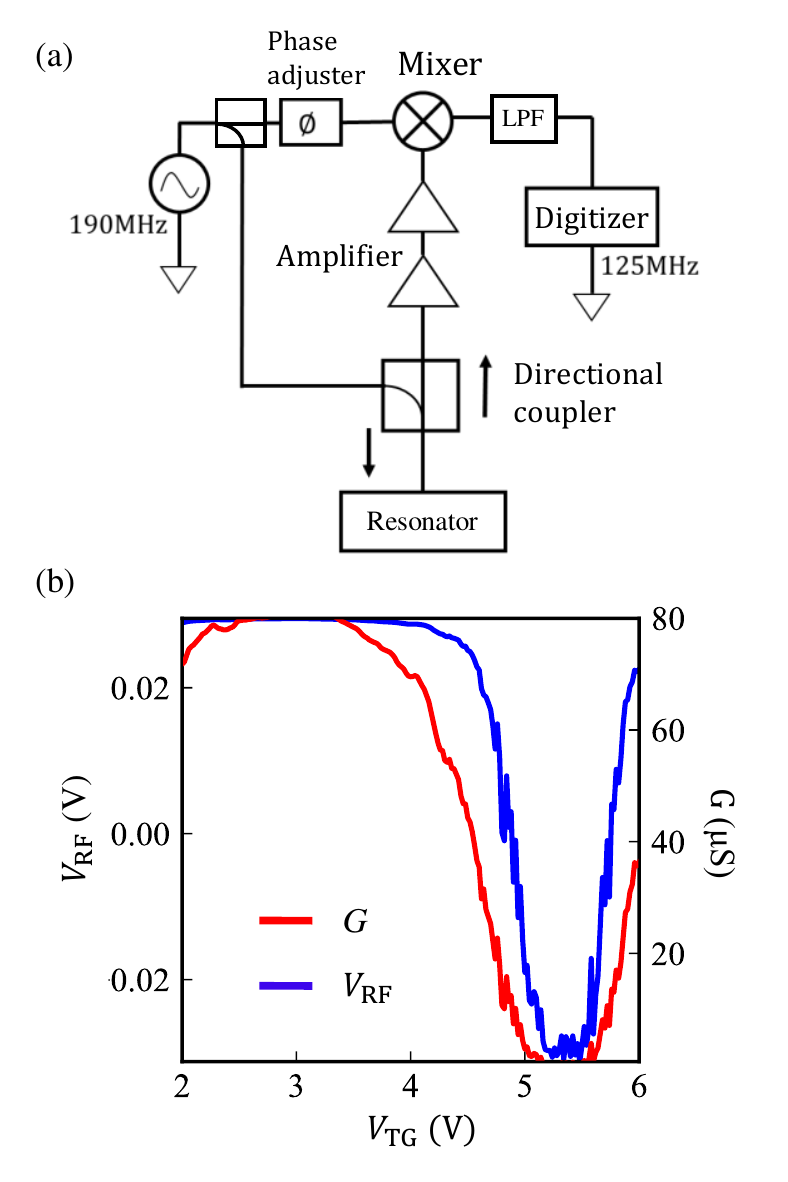}
  \caption{
  (a) Schematic of the demodulator circuit utilizing a mixer for RF-reflectometry. (b) Observed readout signal $V_{\rm RF}$ and the DC conductance $G$ as a function of $V_{\rm TG}$.
  }
  \label{reflectometry}
\end{center}
\end{figure}

We construct a demodulator circuit utilizing a mixer shown in Fig.~\ref{reflectometry} (a).
The RF signal divided by the splitter is injected into the tank circuit including the sample through the directional coupler. The reflected signal is amplified and multiplied with the phase-adjusted input wave, and the resulting voltage through a low pass filter (1.9~MHz) is measured by a digitizer. 

Figure~\ref{reflectometry} (b) show the observed readout signal $V_{\rm RF}$ and the DC conductance $G$ as a function of $V_{TG}$.
The changes of $V_{\rm RF}$ and $G$ are synchronized and we can detect the change of $G$ by monitoring $V_{\rm RF}$.
Note that the observed nonlinearity between $V_{\rm RF}$ and $G$ reflects the nonlinear relation in the reflection coefficient and $G$.


\begin{figure}
\begin{center}
  \includegraphics{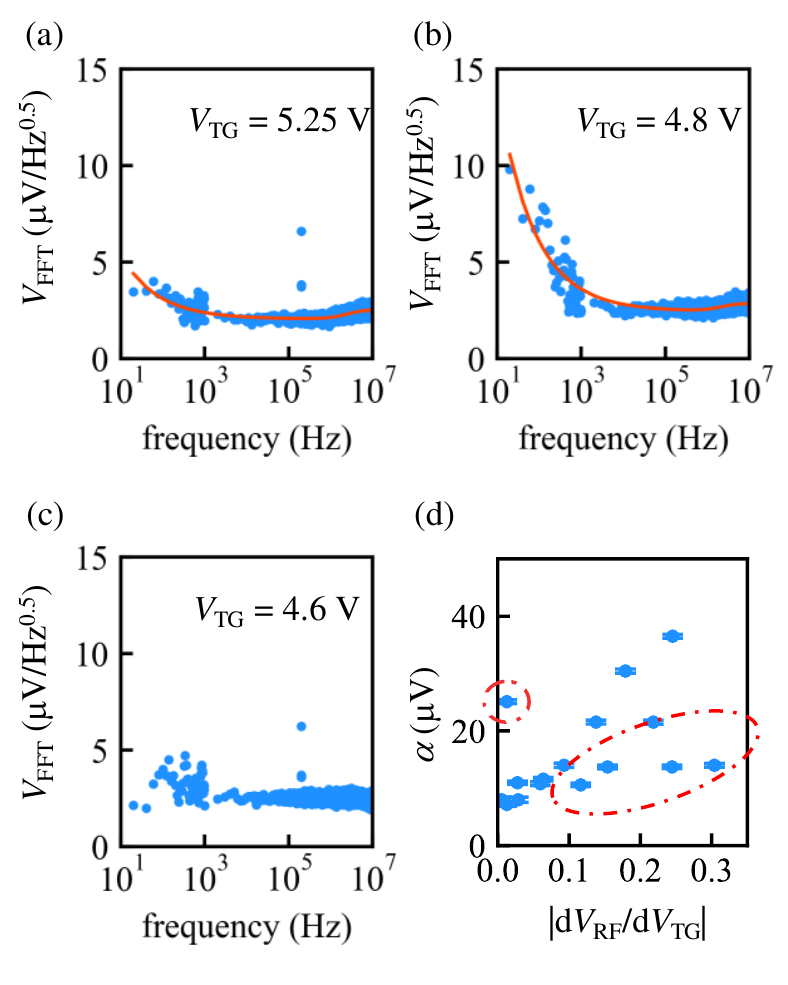}
  \caption{Noise spectra at (a) $V_{\rm TG}$ = 5.25 V, (b) 4.8 V, and (c) 4.6 V. (d) A correlation between $\alpha$ and $\left|\frac{\rm{d} \it{V}_{\rm RF}}{\rm{d} \it{V}_{\rm TG}}\right|$.}
  \label{noise}
\end{center}
\end{figure}

We analyze the readout noise in RF-reflectometry.
Figures~\ref{noise}(a)-(c) show the noise spectrum at $V_{\rm TG}$ = 5.25, 4.8, and 4.6 V, respectively.
The spectra are calculated by fast Fourier transform of the real-time data.
In this measurement, we remove the low pass filter in Fig.~\ref{reflectometry} (a).
At $V_{\rm TG}$ = 5.25 and 4.8 V, we observe the same frequency dependence with RF-reflectometry in GaAs quantum dot~\cite{shinozaki2021gate}.
The solid red lines in Figs~\ref{noise}(a) and (b) are the fitting curve of the following equation.
\begin{align}
    V_{\rm FFT}(f) = \frac{\alpha}{f^{\frac{1}{2}}} - L_{\rm S}(f) + \rm offset
\end{align}
Here, $\alpha$ is the amplitude of the flicker noise corresponding to the device noise, the $L_{\rm S}(f)$ the symmetric Lorentz function to the circuit noise from the characteristics of the resonator and the amplifier, and the offset the intrinsic noise of the amplifier.
On the other hand, such behavior is not observed at 4.6 V in Fig~\ref{noise} (c), indicating that RF-reflectometry is insensitive in this voltage region.
This is because the conductance in such a region makes the resonator impedance far away from the matching conditions and the reflection coefficient is not sensitive to changes of the conductance.
The DC measurement shows that the matching condition is satisfied with $G\sim20~ \rm{\mu} \rm S$ at $V_{\rm TG}\sim 4.8$ V, where the $V_{\rm RF}$ closes to zero.
The reflection coefficient is sensitive enough to detect conductance changes in the pinch-off region, where the conductance becomes lower than $20~ \rm{\mu} \rm S$.
This is important in application to high-sensitive single charge detectors utilizing quantum point contacts and quantum dots, which usually operate at low conductance conditions~\cite{reilly2007fast, 2010BarthelPRB}.

Furthermore, the positive correlation between $\alpha$ and $\left|\frac{\rm{d} \it{V}_{\rm RF}}{\rm{d} \it{V}_{\rm TG}}\right|$ is obtained as shown in Fig.~\ref{noise} (d).
This relationship indicates that the flicker noise is caused by charge fluctuation near the channel in BLG.
Although, there are a few points indicated in the dashed red circle away from the trend.
This is because we calculate $\left|\frac{\rm{d} \it{V}_{\rm RF}}{\rm{d} \it{V}_{\rm TG}}\right|$ by a simple numerical differential of the measured data, which leads to the not accurate values depending on the measurement resolution especially with the fine signal oscillation, such as seen in the Fig.~\ref{reflectometry}(b).
The signal oscillation corresponds to formation of quantum dots and is focused on in a later section.
The present discussion shows that $V_{\rm RF}$ firmly reflects the operation of RF-reflectometory and explains the noise mechanism of this measurement technique.


\begin{figure}
\begin{center}
  \includegraphics{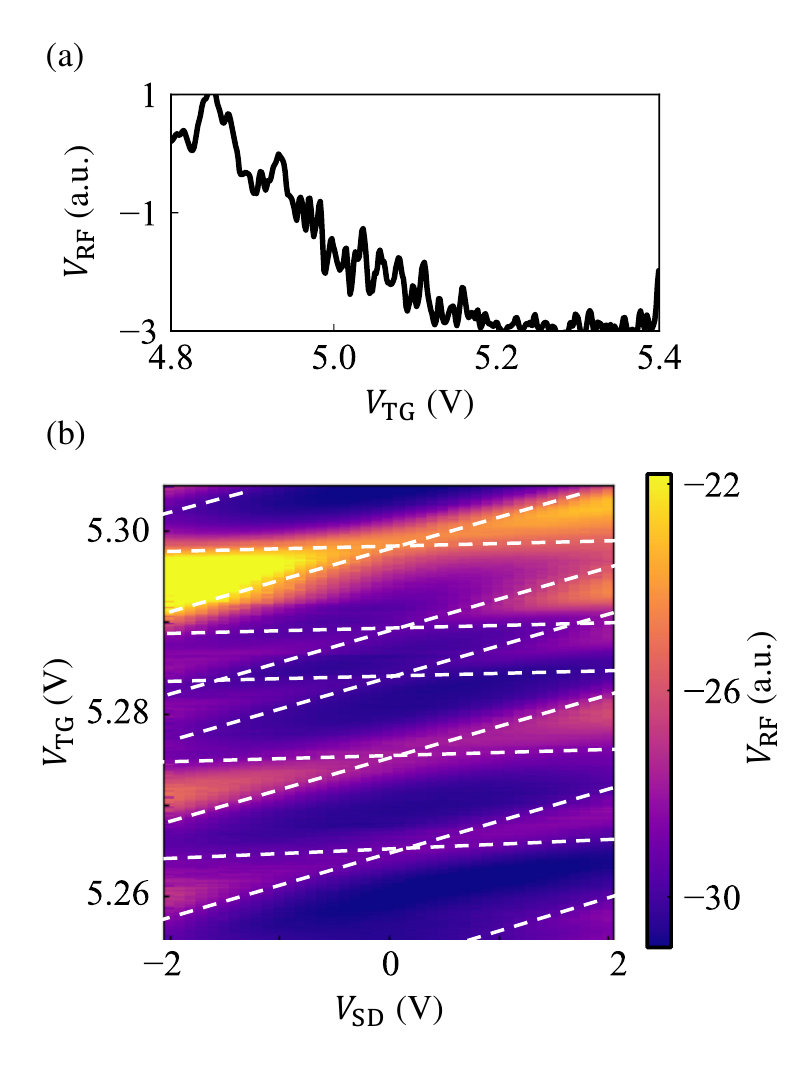}
  \caption{
  (a) Enclosed trace of $V_{\rm RF}$ as a function of $V_{\rm TG}$. (b) Observed $V_{\rm RF}$ as a function of the source drain bias $V_{\rm SD}$ and $V_{\rm TG}$.
  }
  \label{diamond}
\end{center}
\end{figure}

Finally, we use RF-reflectometry to observe the quantum dots unintentionally created in the device.
Figure~\ref{diamond} (a) shows $V_{\rm RF}$ near the pinch-off of the BLG.
Ideally, $V_{\rm RF}$ should show a smooth decrease due to the bandgap caused by the vertical electric field. However, $V_{\rm RF}$ shows oscillations in addition to the background decrease. These oscillations are also observed in DC measurement as shown in Fig.~\ref{reflectometry}(b). 
These are Coulomb peaks.
The observed $V_{\rm RF}$ as a function of the source-drain bias $V_{\rm SD}$ and $V_{\rm TG}$ is shown in Fig.~\ref{diamond} (b).
The signal is also modulated by $V_{\rm SD}$ and Coulomb diamonds are observed (dotted lines).
The possible origin of forming quantum dots will be potential fluctuations due to air bubbles formed when the BLG is encapsulated with the boron nitride.
The bandgap of BLG depends on the vertical electric field. Due to the thickness of the bubbles~\cite{bahamon2015conductance, leconte2017graphene}, the vertical electric field is modulated and quantum dots are created.

\section{Conclusion}
In conclusion, RF-reflectometry in bilayer graphene devices is demonstrated by creating a device with a micro graphite back-gate on an undoped Si substrate. The resonance property of the device connected to the tank circuit is measured using a network analyzer. We confirm that the matching condition close to the operating point can be maintained. We form RF-reflectometory setup and compared the result with the DC measurement, and confirmed their consistency. We also measure Coulomb diamonds of quantum dots possibly formed by bubbles and confirm that RF-reflectometry of quantum dots can be performed. Our technique enables high-speed measurements of bilayer graphene quantum dots devices and helps the research of bilayer graphene based quantum bits by fast readout of the states.

\section{Acknowledgements}
The authors thank S. Masubuchi, T. Machida, S. Alka, T. Kitada, T. Kumasaka, and RIEC Fundamental Technology Center and the Laboratory for Nanoelectronics and Spintronics for fruitful discussion and technical support.
Part of this work is supported by 
MEXT Leading Initiative for Excellent Young Researchers, 
Grants-in-Aid for Scientific Research (20H00237, 21K18592), 
Fujikura Foundation Research Grant, 
Hattori Hokokai Foudation Research Grant, 
Kondo Zaidan Research Grant, 
Tanigawa Foundation Research Grant, 
FRiD Tohoku University. 

\section{Author contributions}
T. O. planned the project;
T. J. and T. O. performed device fabrication; 
T. J., Y. F., T. A., and T. O. developed the measurement setup;
T. J., M. S., and T. O. conducted experiments and data analysis; all authors discussed the results;
T. J., M. S., and T. O. wrote the manuscript with input from all the authors.

\bibliography{reference}

\providecommand{\noopsort}[1]{}\providecommand{\singleletter}[1]{#1}%
\providecommand{\latin}[1]{#1}
\makeatletter
\providecommand{\doi}
  {\begingroup\let\do\@makeother\dospecials
  \catcode`\{=1 \catcode`\}=2 \doi@aux}
\providecommand{\doi@aux}[1]{\endgroup\texttt{#1}}
\makeatother
\providecommand*\mcitethebibliography{\thebibliography}
\csname @ifundefined\endcsname{endmcitethebibliography}
  {\let\endmcitethebibliography\endthebibliography}{}
\begin{mcitethebibliography}{43}
\providecommand*\natexlab[1]{#1}
\providecommand*\mciteSetBstSublistMode[1]{}
\providecommand*\mciteSetBstMaxWidthForm[2]{}
\providecommand*\mciteBstWouldAddEndPuncttrue
  {\def\EndOfBibitem{\unskip.}}
\providecommand*\mciteBstWouldAddEndPunctfalse
  {\let\EndOfBibitem\relax}
\providecommand*\mciteSetBstMidEndSepPunct[3]{}
\providecommand*\mciteSetBstSublistLabelBeginEnd[3]{}
\providecommand*\EndOfBibitem{}
\mciteSetBstSublistMode{f}
\mciteSetBstMaxWidthForm{subitem}{(\alph{mcitesubitemcount})}
\mciteSetBstSublistLabelBeginEnd
  {\mcitemaxwidthsubitemform\space}
  {\relax}
  {\relax}

\bibitem[Zhang \latin{et~al.}(2009)Zhang, Tang, Girit, Hao, Martin, Zettl,
  Crommie, Shen, and Wang]{zhang2009direct}
Zhang,~Y.; Tang,~T.-T.; Girit,~C.; Hao,~Z.; Martin,~M.~C.; Zettl,~A.;
  Crommie,~M.~F.; Shen,~Y.~R.; Wang,~F. Direct observation of a widely tunable
  bandgap in bilayer graphene. \emph{Nature} \textbf{2009}, \emph{459},
  820--823\relax
\mciteBstWouldAddEndPuncttrue
\mciteSetBstMidEndSepPunct{\mcitedefaultmidpunct}
{\mcitedefaultendpunct}{\mcitedefaultseppunct}\relax
\EndOfBibitem
\bibitem[Dean \latin{et~al.}(2010)Dean, Young, Meric, Lee, Wang, Sorgenfrei,
  Watanabe, Taniguchi, Kim, Shepard, \latin{et~al.} others]{dean2010boron}
Dean,~C.~R.; Young,~A.~F.; Meric,~I.; Lee,~C.; Wang,~L.; Sorgenfrei,~S.;
  Watanabe,~K.; Taniguchi,~T.; Kim,~P.; Shepard,~K.~L., \latin{et~al.}  Boron
  nitride substrates for high-quality graphene electronics. \emph{Nat.
  Nanotechnol.} \textbf{2010}, \emph{5}, 722--726\relax
\mciteBstWouldAddEndPuncttrue
\mciteSetBstMidEndSepPunct{\mcitedefaultmidpunct}
{\mcitedefaultendpunct}{\mcitedefaultseppunct}\relax
\EndOfBibitem
\bibitem[Moriyama \latin{et~al.}(2009)Moriyama, Tsuya, Watanabe, Uji, Shimizu,
  Mori, Yamaguchi, and Ishibashi]{2009MoriyamaNanoLett}
Moriyama,~S.; Tsuya,~D.; Watanabe,~E.; Uji,~S.; Shimizu,~M.; Mori,~T.;
  Yamaguchi,~T.; Ishibashi,~K. Coupled Quantum Dots in a Graphene-Based
  Two-Dimensional Semimetal. \emph{Nano Lett.} \textbf{2009}, \emph{9},
  2891--2896\relax
\mciteBstWouldAddEndPuncttrue
\mciteSetBstMidEndSepPunct{\mcitedefaultmidpunct}
{\mcitedefaultendpunct}{\mcitedefaultseppunct}\relax
\EndOfBibitem
\bibitem[G\"{u}ttinger \latin{et~al.}(2011)G\"{u}ttinger, Seif, Stampfer,
  Capelli, Ensslin, and Ihn]{2011GutingerPRB}
G\"{u}ttinger,~J.; Seif,~J.; Stampfer,~C.; Capelli,~A.; Ensslin,~K.; Ihn,~T.
  Time-resolved charge detection in graphene quantum dots. \emph{Phys. Rev. B}
  \textbf{2011}, \emph{83}, 165445\relax
\mciteBstWouldAddEndPuncttrue
\mciteSetBstMidEndSepPunct{\mcitedefaultmidpunct}
{\mcitedefaultendpunct}{\mcitedefaultseppunct}\relax
\EndOfBibitem
\bibitem[Volk \latin{et~al.}(2011)Volk, Fringes, Terrés, Dauber, Engels,
  Trellenkamp, and Stampfer]{2011VolkNanoLett}
Volk,~C.; Fringes,~S.; Terrés,~B.; Dauber,~J.; Engels,~S.; Trellenkamp,~S.;
  Stampfer,~C. Electronic Excited States in Bilayer Graphene Double Quantum
  Dots. \emph{Nano Lett.} \textbf{2011}, \emph{11}, 3581–3586\relax
\mciteBstWouldAddEndPuncttrue
\mciteSetBstMidEndSepPunct{\mcitedefaultmidpunct}
{\mcitedefaultendpunct}{\mcitedefaultseppunct}\relax
\EndOfBibitem
\bibitem[Eich \latin{et~al.}(2018)Eich, Herman, Pisoni, Overweg, Kurzmann, Lee,
  Rickhaus, Watanabe, Taniguchi, Sigrist, Ihn, and Ensslin]{2018EichPRX}
Eich,~M.; Herman,~F.; Pisoni,~R.; Overweg,~H.; Kurzmann,~A.; Lee,~Y.;
  Rickhaus,~P.; Watanabe,~K.; Taniguchi,~T.; Sigrist,~M.; Ihn,~T.; Ensslin,~K.
  Spin and Valley States in Gate-Defined Bilayer Graphene Quantum Dots.
  \emph{Phys. Rev. X} \textbf{2018}, \emph{8}, 031023\relax
\mciteBstWouldAddEndPuncttrue
\mciteSetBstMidEndSepPunct{\mcitedefaultmidpunct}
{\mcitedefaultendpunct}{\mcitedefaultseppunct}\relax
\EndOfBibitem
\bibitem[Kurzmann \latin{et~al.}(2019)Kurzmann, Overweg, Eich, Pally, Rickhaus,
  Pisoni, Lee, Watanabe, Taniguchi, Ihn, \latin{et~al.}
  others]{kurzmann2019charge}
Kurzmann,~A.; Overweg,~H.; Eich,~M.; Pally,~A.; Rickhaus,~P.; Pisoni,~R.;
  Lee,~Y.; Watanabe,~K.; Taniguchi,~T.; Ihn,~T., \latin{et~al.}  Charge
  detection in gate-defined bilayer graphene quantum dots. \emph{Nano Lett.}
  \textbf{2019}, \emph{19}, 5216--5221\relax
\mciteBstWouldAddEndPuncttrue
\mciteSetBstMidEndSepPunct{\mcitedefaultmidpunct}
{\mcitedefaultendpunct}{\mcitedefaultseppunct}\relax
\EndOfBibitem
\bibitem[Kurzmann \latin{et~al.}(2019)Kurzmann, Eich, Overweg, Mangold, Herman,
  Rickhaus, Pisoni, Lee, Garreis, Tong, \latin{et~al.}
  others]{kurzmann2019excited}
Kurzmann,~A.; Eich,~M.; Overweg,~H.; Mangold,~M.; Herman,~F.; Rickhaus,~P.;
  Pisoni,~R.; Lee,~Y.; Garreis,~R.; Tong,~C., \latin{et~al.}  Excited states in
  bilayer graphene quantum dots. \emph{Phys. Rev. Lett.} \textbf{2019},
  \emph{123}, 026803\relax
\mciteBstWouldAddEndPuncttrue
\mciteSetBstMidEndSepPunct{\mcitedefaultmidpunct}
{\mcitedefaultendpunct}{\mcitedefaultseppunct}\relax
\EndOfBibitem
\bibitem[Banszerus \latin{et~al.}(2020)Banszerus, M\"{o}ller, Icking, Watanabe,
  Taniguchi, Volk, and Stampfer]{banszerus2020single}
Banszerus,~L.; M\"{o}ller,~S.; Icking,~E.; Watanabe,~K.; Taniguchi,~T.;
  Volk,~C.; Stampfer,~C. Single-electron double quantum dots in bilayer
  graphene. \emph{Nano Lett.} \textbf{2020}, \emph{20}, 2005--2011\relax
\mciteBstWouldAddEndPuncttrue
\mciteSetBstMidEndSepPunct{\mcitedefaultmidpunct}
{\mcitedefaultendpunct}{\mcitedefaultseppunct}\relax
\EndOfBibitem
\bibitem[Banszerus \latin{et~al.}(2021)Banszerus, Hecker, Icking, Trellenkamp,
  Lentz, Neumaier, Watanabe, Taniguchi, Volk, and
  Stampfer]{banszerus2021pulsed}
Banszerus,~L.; Hecker,~K.; Icking,~E.; Trellenkamp,~S.; Lentz,~F.;
  Neumaier,~D.; Watanabe,~K.; Taniguchi,~T.; Volk,~C.; Stampfer,~C. Pulsed-gate
  spectroscopy of single-electron spin states in bilayer graphene quantum dots.
  \emph{Phys. Rev. B} \textbf{2021}, \emph{103}, L081404\relax
\mciteBstWouldAddEndPuncttrue
\mciteSetBstMidEndSepPunct{\mcitedefaultmidpunct}
{\mcitedefaultendpunct}{\mcitedefaultseppunct}\relax
\EndOfBibitem
\bibitem[Banszerus \latin{et~al.}(2021)Banszerus, M\"{o}ller, Steiner, Icking,
  Trellenkamp, Lentz, Watanabe, Taniguchi, Volk, and
  Stampfer]{banszerus2021spin}
Banszerus,~L.; M\"{o}ller,~S.; Steiner,~C.; Icking,~E.; Trellenkamp,~S.;
  Lentz,~F.; Watanabe,~K.; Taniguchi,~T.; Volk,~C.; Stampfer,~C. Spin-valley
  coupling in single-electron bilayer graphene quantum dots. \emph{Nat.
  Commun.} \textbf{2021}, \emph{12}, 1--7\relax
\mciteBstWouldAddEndPuncttrue
\mciteSetBstMidEndSepPunct{\mcitedefaultmidpunct}
{\mcitedefaultendpunct}{\mcitedefaultseppunct}\relax
\EndOfBibitem
\bibitem[Tarucha \latin{et~al.}(1996)Tarucha, Austing, Honda, Van~der Hage, and
  Kouwenhoven]{tarucha1996shell}
Tarucha,~S.; Austing,~D.; Honda,~T.; Van~der Hage,~R.; Kouwenhoven,~L.~P. Shell
  filling and spin effects in a few electron quantum dot. \emph{Phys. Rev.
  Lett.} \textbf{1996}, \emph{77}, 3613\relax
\mciteBstWouldAddEndPuncttrue
\mciteSetBstMidEndSepPunct{\mcitedefaultmidpunct}
{\mcitedefaultendpunct}{\mcitedefaultseppunct}\relax
\EndOfBibitem
\bibitem[Kouwenhoven \latin{et~al.}(1997)Kouwenhoven, Oosterkamp, Danoesastro,
  Eto, Austing, Honda, and Tarucha]{kouwenhoven1997excitation}
Kouwenhoven,~L.~P.; Oosterkamp,~T.; Danoesastro,~M.; Eto,~M.; Austing,~D.;
  Honda,~T.; Tarucha,~S. Excitation spectra of circular, few-electron quantum
  dots. \emph{Science} \textbf{1997}, \emph{278}, 1788--1792\relax
\mciteBstWouldAddEndPuncttrue
\mciteSetBstMidEndSepPunct{\mcitedefaultmidpunct}
{\mcitedefaultendpunct}{\mcitedefaultseppunct}\relax
\EndOfBibitem
\bibitem[Kouwenhoven \latin{et~al.}(2001)Kouwenhoven, Austing, and
  Tarucha]{kouwenhoven2001few}
Kouwenhoven,~L.~P.; Austing,~D.; Tarucha,~S. Few-electron quantum dots.
  \emph{Rep. Prog. Phys.} \textbf{2001}, \emph{64}, 701\relax
\mciteBstWouldAddEndPuncttrue
\mciteSetBstMidEndSepPunct{\mcitedefaultmidpunct}
{\mcitedefaultendpunct}{\mcitedefaultseppunct}\relax
\EndOfBibitem
\bibitem[Hanson \latin{et~al.}(2007)Hanson, Kouwenhoven, Petta, Tarucha, and
  Vandersypen]{2007HansonRMP}
Hanson,~R.; Kouwenhoven,~L.~P.; Petta,~J.~R.; Tarucha,~S.; Vandersypen,~L.
  M.~K. Spins in few-electron quantum dots. \emph{Rev. Mod. Phys.}
  \textbf{2007}, \emph{79}, 1217\relax
\mciteBstWouldAddEndPuncttrue
\mciteSetBstMidEndSepPunct{\mcitedefaultmidpunct}
{\mcitedefaultendpunct}{\mcitedefaultseppunct}\relax
\EndOfBibitem
\bibitem[Loss and DiVincenzo(1998)Loss, and DiVincenzo]{1998LossPRA}
Loss,~D.; DiVincenzo,~D.~P. Quantum computation with quantum dots. \emph{Phys.
  Rev. A.} \textbf{1998}, \emph{57}, 120\relax
\mciteBstWouldAddEndPuncttrue
\mciteSetBstMidEndSepPunct{\mcitedefaultmidpunct}
{\mcitedefaultendpunct}{\mcitedefaultseppunct}\relax
\EndOfBibitem
\bibitem[Petta \latin{et~al.}(2005)Petta, Johnson, Taylor, Laird, Yacoby,
  Lukin, Marcus, Hanson, and Gossard]{2005PettaSci}
Petta,~J.~R.; Johnson,~A.~C.; Taylor,~J.~M.; Laird,~E.~A.; Yacoby,~A.;
  Lukin,~M.~D.; Marcus,~C.~M.; Hanson,~M.~P.; Gossard,~A.~C. Coherent
  Manipulation of Coupled Electron Spins in Semiconductor Quantum Dots.
  \emph{Science} \textbf{2005}, \emph{309}, 2180--2184\relax
\mciteBstWouldAddEndPuncttrue
\mciteSetBstMidEndSepPunct{\mcitedefaultmidpunct}
{\mcitedefaultendpunct}{\mcitedefaultseppunct}\relax
\EndOfBibitem
\bibitem[Koppens \latin{et~al.}(2006)Koppens, Buizert, Tielrooij, Vink, Nowack,
  Meunier, Kouwenhoven, and Vandersypen]{2006KoppensNature}
Koppens,~F. H.~L.; Buizert,~C.; Tielrooij,~K.~J.; Vink,~I.~T.; Nowack,~K.~C.;
  Meunier,~T.; Kouwenhoven,~L.~P.; Vandersypen,~L. M.~K. Driven coherent
  oscillations of a single electron spin in a quantum dot. \emph{Nature}
  \textbf{2006}, \emph{442}, 766--771\relax
\mciteBstWouldAddEndPuncttrue
\mciteSetBstMidEndSepPunct{\mcitedefaultmidpunct}
{\mcitedefaultendpunct}{\mcitedefaultseppunct}\relax
\EndOfBibitem
\bibitem[Ladd \latin{et~al.}(2010)Ladd, Jelezko, Laflamme, Nakamura, Monroe,
  and O’Brien]{2010LaddNature}
Ladd,~T.~D.; Jelezko,~F.; Laflamme,~R.; Nakamura,~Y.; Monroe,~C.;
  O’Brien,~J.~L. Quantum computers. \emph{Nature} \textbf{2010}, \emph{464},
  45--53\relax
\mciteBstWouldAddEndPuncttrue
\mciteSetBstMidEndSepPunct{\mcitedefaultmidpunct}
{\mcitedefaultendpunct}{\mcitedefaultseppunct}\relax
\EndOfBibitem
\bibitem[Veldhorst \latin{et~al.}(2015)Veldhorst, Yang, Hwang, Huang,
  Dehollain, Muhonen, Simmons, Laucht, Hudson, Itoh, Morello, and
  Dzurak]{2015VeldhorstNat}
Veldhorst,~M.; Yang,~C.~H.; Hwang,~J. C.~C.; Huang,~W.; Dehollain,~J.~P.;
  Muhonen,~J.~T.; Simmons,~S.; Laucht,~A.; Hudson,~F.~E.; Itoh,~K.~M.;
  Morello,~A.; Dzurak,~A.~S. A two-qubit logic gate in silicon. \emph{Nature}
  \textbf{2015}, \emph{526}, 410--414\relax
\mciteBstWouldAddEndPuncttrue
\mciteSetBstMidEndSepPunct{\mcitedefaultmidpunct}
{\mcitedefaultendpunct}{\mcitedefaultseppunct}\relax
\EndOfBibitem
\bibitem[Otsuka \latin{et~al.}(2016)Otsuka, Nakajima, Delbecq, Amaha, Yoneda,
  Takeda, Allison, Ito, Sugawara, Noiri, Ludwig, Wieck, and
  Tarucha]{2016OtsukaSciRep}
Otsuka,~T.; Nakajima,~T.; Delbecq,~M.~R.; Amaha,~S.; Yoneda,~J.; Takeda,~K.;
  Allison,~G.; Ito,~T.; Sugawara,~R.; Noiri,~A.; Ludwig,~A.; Wieck,~A.~D.;
  Tarucha,~S. Single-electron Spin Resonance in a Quadruple Quantum Dot.
  \emph{Sci. Rep.} \textbf{2016}, \emph{6}, 31820\relax
\mciteBstWouldAddEndPuncttrue
\mciteSetBstMidEndSepPunct{\mcitedefaultmidpunct}
{\mcitedefaultendpunct}{\mcitedefaultseppunct}\relax
\EndOfBibitem
\bibitem[Yoneda \latin{et~al.}(2018)Yoneda, Takeda, Otsuka, Nakajima, Delbecq,
  Allison, Honda, Kodera, Oda, Hoshi, Usami, Itoh, and
  Tarucha]{2018YonedaNatNano}
Yoneda,~J.; Takeda,~K.; Otsuka,~T.; Nakajima,~T.; Delbecq,~M.~R.; Allison,~G.;
  Honda,~T.; Kodera,~T.; Oda,~S.; Hoshi,~Y.; Usami,~N.; Itoh,~K.~M.;
  Tarucha,~S. A quantum-dot spin qubit with coherence limited by charge noise
  and fidelity higher than 99.9\%. \emph{Nat. Nanotechnol.} \textbf{2018},
  \emph{13}, 102--106\relax
\mciteBstWouldAddEndPuncttrue
\mciteSetBstMidEndSepPunct{\mcitedefaultmidpunct}
{\mcitedefaultendpunct}{\mcitedefaultseppunct}\relax
\EndOfBibitem
\bibitem[Altimiras \latin{et~al.}(2010)Altimiras, le~Sueur, Gennser, Cavanna,
  Mailly, and Pierre]{2010AltimirasPRL}
Altimiras,~C.; le~Sueur,~H.; Gennser,~U.; Cavanna,~A.; Mailly,~D.; Pierre,~F.
  Tuning energy relaxation along quantum hall channels. \emph{Phys. Rev. Lett.}
  \textbf{2010}, \emph{105}, 226804\relax
\mciteBstWouldAddEndPuncttrue
\mciteSetBstMidEndSepPunct{\mcitedefaultmidpunct}
{\mcitedefaultendpunct}{\mcitedefaultseppunct}\relax
\EndOfBibitem
\bibitem[Otsuka \latin{et~al.}(2015)Otsuka, Amaha, Nakajima, Delbecq, Yoneda,
  Takeda, Sugawara, Allison, Ludwig, Wieck, and Tarucha]{2015OtsukaSciRep}
Otsuka,~T.; Amaha,~S.; Nakajima,~T.; Delbecq,~M.~R.; Yoneda,~J.; Takeda,~K.;
  Sugawara,~R.; Allison,~G.; Ludwig,~A.; Wieck,~A.~D.; Tarucha,~S. Fast probe
  of local electronic states in nanostructures utilizing a single-lead quantum
  dot. \emph{Sci. Pep.} \textbf{2015}, \emph{5}, 14616\relax
\mciteBstWouldAddEndPuncttrue
\mciteSetBstMidEndSepPunct{\mcitedefaultmidpunct}
{\mcitedefaultendpunct}{\mcitedefaultseppunct}\relax
\EndOfBibitem
\bibitem[Otsuka \latin{et~al.}(2019)Otsuka, Nakajima, Delbecq, Stano, Amaha,
  Yoneda, Takeda, Allison, Li, Noiri, Ito, Loss, Ludwig, Wieck, and
  Tarucha]{2019OtsukaPRB}
Otsuka,~T.; Nakajima,~T.; Delbecq,~M.~R.; Stano,~P.; Amaha,~S.; Yoneda,~J.;
  Takeda,~K.; Allison,~G.; Li,~S.; Noiri,~A.; Ito,~T.; Loss,~D.; Ludwig,~A.;
  Wieck,~A.~D.; Tarucha,~S. Difference in charge and spin dynamics in a quantum
  dot–lead coupled system. \emph{Phys. Rev. B} \textbf{2019}, \emph{99},
  085402\relax
\mciteBstWouldAddEndPuncttrue
\mciteSetBstMidEndSepPunct{\mcitedefaultmidpunct}
{\mcitedefaultendpunct}{\mcitedefaultseppunct}\relax
\EndOfBibitem
\bibitem[Trauzettel \latin{et~al.}(2007)Trauzettel, Bulaev, Loss, and
  Burkard]{2007TrauzettelNatPhys}
Trauzettel,~B.; Bulaev,~D.~V.; Loss,~D.; Burkard,~G. Spin qubits in graphene
  quantum dots. \emph{Nat. Phys.} \textbf{2007}, \emph{3}, 192\relax
\mciteBstWouldAddEndPuncttrue
\mciteSetBstMidEndSepPunct{\mcitedefaultmidpunct}
{\mcitedefaultendpunct}{\mcitedefaultseppunct}\relax
\EndOfBibitem
\bibitem[Schoelkopf \latin{et~al.}(1998)Schoelkopf, Wahlgren, Kozhevnikov,
  Delsing, and Prober]{schoelkopf1998radio}
Schoelkopf,~R.; Wahlgren,~P.; Kozhevnikov,~A.; Delsing,~P.; Prober,~D. The
  radio-frequency single-electron transistor (RF-SET): A fast and
  ultrasensitive electrometer. \emph{Science} \textbf{1998}, \emph{280},
  1238--1242\relax
\mciteBstWouldAddEndPuncttrue
\mciteSetBstMidEndSepPunct{\mcitedefaultmidpunct}
{\mcitedefaultendpunct}{\mcitedefaultseppunct}\relax
\EndOfBibitem
\bibitem[Qin and Williams(2006)Qin, and Williams]{qin2006radio}
Qin,~H.; Williams,~D.~A. Radio-frequency point-contact electrometer.
  \emph{Appl. Phys. Lett.} \textbf{2006}, \emph{88}, 203506\relax
\mciteBstWouldAddEndPuncttrue
\mciteSetBstMidEndSepPunct{\mcitedefaultmidpunct}
{\mcitedefaultendpunct}{\mcitedefaultseppunct}\relax
\EndOfBibitem
\bibitem[Reilly \latin{et~al.}(2007)Reilly, Marcus, Hanson, and
  Gossard]{reilly2007fast}
Reilly,~D.; Marcus,~C.; Hanson,~M.; Gossard,~A. Fast single-charge sensing with
  a rf quantum point contact. \emph{Appl. Phys. Lett.} \textbf{2007},
  \emph{91}, 162101\relax
\mciteBstWouldAddEndPuncttrue
\mciteSetBstMidEndSepPunct{\mcitedefaultmidpunct}
{\mcitedefaultendpunct}{\mcitedefaultseppunct}\relax
\EndOfBibitem
\bibitem[Barthel \latin{et~al.}(2009)Barthel, Reilly, Marcus, Hanson, and
  Gossard]{barthel2009rapid}
Barthel,~C.; Reilly,~D.; Marcus,~C.~M.; Hanson,~M.; Gossard,~A. Rapid
  single-shot measurement of a singlet-triplet qubit. \emph{Phys. Rev. Lett.}
  \textbf{2009}, \emph{103}, 160503\relax
\mciteBstWouldAddEndPuncttrue
\mciteSetBstMidEndSepPunct{\mcitedefaultmidpunct}
{\mcitedefaultendpunct}{\mcitedefaultseppunct}\relax
\EndOfBibitem
\bibitem[Barthel \latin{et~al.}(2010)Barthel, Kj\ae~rgaard, Medford, Stopa,
  Marcus, Hanson, and Gossard]{2010BarthelPRB}
Barthel,~C.; Kj\ae~rgaard,~M.; Medford,~J.; Stopa,~M.; Marcus,~C.~M.;
  Hanson,~M.~P.; Gossard,~A.~C. Fast sensing of double-dot charge arrangement
  and spin state with a radio-frequency sensor quantum dot. \emph{Phys. Rev. B}
  \textbf{2010}, \emph{81}, 161308\relax
\mciteBstWouldAddEndPuncttrue
\mciteSetBstMidEndSepPunct{\mcitedefaultmidpunct}
{\mcitedefaultendpunct}{\mcitedefaultseppunct}\relax
\EndOfBibitem
\bibitem[Banszerus \latin{et~al.}(2021)Banszerus, M{\"o}ller, Icking, Steiner,
  Neumaier, Otto, Watanabe, Taniguchi, Volk, and
  Stampfer]{banszerus2021dispersive}
Banszerus,~L.; M{\"o}ller,~S.; Icking,~E.; Steiner,~C.; Neumaier,~D.; Otto,~M.;
  Watanabe,~K.; Taniguchi,~T.; Volk,~C.; Stampfer,~C. Dispersive sensing of
  charge states in a bilayer graphene quantum dot. \emph{Appl. Phys. Lett.}
  \textbf{2021}, \emph{118}, 093104\relax
\mciteBstWouldAddEndPuncttrue
\mciteSetBstMidEndSepPunct{\mcitedefaultmidpunct}
{\mcitedefaultendpunct}{\mcitedefaultseppunct}\relax
\EndOfBibitem
\bibitem[Fu \latin{et~al.}(2014)Fu, El~Abbassi, Hasler, Jung, Steinacher,
  Calame, Sch{\"o}nenberger, Puebla-Hellmann, Hellm{\"u}ller, Ihn,
  \latin{et~al.} others]{fu2014electrolyte}
Fu,~W.; El~Abbassi,~M.; Hasler,~T.; Jung,~M.; Steinacher,~M.; Calame,~M.;
  Sch{\"o}nenberger,~C.; Puebla-Hellmann,~G.; Hellm{\"u}ller,~S.; Ihn,~T.,
  \latin{et~al.}  Electrolyte gate dependent high-frequency measurement of
  graphene field-effect transistor for sensing applications. \emph{Appl. Phys.
  Lett.} \textbf{2014}, \emph{104}, 013102\relax
\mciteBstWouldAddEndPuncttrue
\mciteSetBstMidEndSepPunct{\mcitedefaultmidpunct}
{\mcitedefaultendpunct}{\mcitedefaultseppunct}\relax
\EndOfBibitem
\bibitem[M{\"u}ller \latin{et~al.}(2012)M{\"u}ller, G{\"u}ttinger, Bischoff,
  Hellm{\"u}ller, Ensslin, and Ihn]{muller2012fast}
M{\"u}ller,~T.; G{\"u}ttinger,~J.; Bischoff,~D.; Hellm{\"u}ller,~S.;
  Ensslin,~K.; Ihn,~T. Fast detection of single-charge tunneling to a graphene
  quantum dot in a multi-level regime. \emph{Appl. Phys. Lett.} \textbf{2012},
  \emph{101}, 012104\relax
\mciteBstWouldAddEndPuncttrue
\mciteSetBstMidEndSepPunct{\mcitedefaultmidpunct}
{\mcitedefaultendpunct}{\mcitedefaultseppunct}\relax
\EndOfBibitem
\bibitem[Noiri \latin{et~al.}(2020)Noiri, Takeda, Yoneda, Nakajima, Kodera, and
  Tarucha]{noiri2020radio}
Noiri,~A.; Takeda,~K.; Yoneda,~J.; Nakajima,~T.; Kodera,~T.; Tarucha,~S.
  Radio-frequency-detected fast charge sensing in undoped silicon quantum dots.
  \emph{Nano Lett.} \textbf{2020}, \emph{20}, 947--952\relax
\mciteBstWouldAddEndPuncttrue
\mciteSetBstMidEndSepPunct{\mcitedefaultmidpunct}
{\mcitedefaultendpunct}{\mcitedefaultseppunct}\relax
\EndOfBibitem
\bibitem[Mizokuchi \latin{et~al.}(2021)Mizokuchi, Bugu, Hirayama, Yoneda, and
  Kodera]{mizokuchi2021radio}
Mizokuchi,~R.; Bugu,~S.; Hirayama,~M.; Yoneda,~J.; Kodera,~T. Radio-frequency
  single electron transistors in physically defined silicon quantum dots with a
  sensitive phase response. \emph{Sci. Rep.} \textbf{2021}, \emph{11},
  1--7\relax
\mciteBstWouldAddEndPuncttrue
\mciteSetBstMidEndSepPunct{\mcitedefaultmidpunct}
{\mcitedefaultendpunct}{\mcitedefaultseppunct}\relax
\EndOfBibitem
\bibitem[Bugu \latin{et~al.}(2021)Bugu, Nishiyama, Kato, Liu, Mori, and
  Kodera]{2021BuguJJAP}
Bugu,~S.; Nishiyama,~S.; Kato,~K.; Liu,~Y.; Mori,~T.; Kodera,~T. RF
  reflectometry for readout of charge transition in a physically defined
  p-channel MOS silicon quantum dot. \emph{Jpn. J. Appl. Phys.} \textbf{2021},
  \emph{60}, 1--4\relax
\mciteBstWouldAddEndPuncttrue
\mciteSetBstMidEndSepPunct{\mcitedefaultmidpunct}
{\mcitedefaultendpunct}{\mcitedefaultseppunct}\relax
\EndOfBibitem
\bibitem[Bahamon \latin{et~al.}(2015)Bahamon, Qi, Park, Pereira, and
  Campbell]{bahamon2015conductance}
Bahamon,~D.~A.; Qi,~Z.; Park,~H.~S.; Pereira,~V.~M.; Campbell,~D.~K.
  Conductance signatures of electron confinement induced by strained
  nanobubbles in graphene. \emph{Nanoscale} \textbf{2015}, \emph{7},
  15300--15309\relax
\mciteBstWouldAddEndPuncttrue
\mciteSetBstMidEndSepPunct{\mcitedefaultmidpunct}
{\mcitedefaultendpunct}{\mcitedefaultseppunct}\relax
\EndOfBibitem
\bibitem[Leconte \latin{et~al.}(2017)Leconte, Kim, Kim, Ha, Watanabe,
  Taniguchi, Jung, and Jung]{leconte2017graphene}
Leconte,~N.; Kim,~H.; Kim,~H.-J.; Ha,~D.~H.; Watanabe,~K.; Taniguchi,~T.;
  Jung,~J.; Jung,~S. Graphene bubbles and their role in graphene quantum
  transport. \emph{Nanoscale} \textbf{2017}, \emph{9}, 6041--6047\relax
\mciteBstWouldAddEndPuncttrue
\mciteSetBstMidEndSepPunct{\mcitedefaultmidpunct}
{\mcitedefaultendpunct}{\mcitedefaultseppunct}\relax
\EndOfBibitem
\bibitem[Bertolazzi \latin{et~al.}(2019)Bertolazzi, Bondavalli, Roche, San,
  Choi, Colombo, Bonaccorso, and Samori]{bertolazzi2019nonvolatile}
Bertolazzi,~S.; Bondavalli,~P.; Roche,~S.; San,~T.; Choi,~S.-Y.; Colombo,~L.;
  Bonaccorso,~F.; Samori,~P. Nonvolatile memories based on graphene and related
  2D materials. \emph{Adv. Mater.} \textbf{2019}, \emph{31}, 1806663\relax
\mciteBstWouldAddEndPuncttrue
\mciteSetBstMidEndSepPunct{\mcitedefaultmidpunct}
{\mcitedefaultendpunct}{\mcitedefaultseppunct}\relax
\EndOfBibitem
\bibitem[Masubuchi \latin{et~al.}(2018)Masubuchi, Morimoto, Morikawa, Onodera,
  Asakawa, Watanabe, Taniguchi, and Machida]{2018MasubuchiNatCom}
Masubuchi,~S.; Morimoto,~M.; Morikawa,~S.; Onodera,~M.; Asakawa,~Y.;
  Watanabe,~K.; Taniguchi,~T.; Machida,~T. Autonomous robotic searching and
  assembly of two-dimensional crystals to build van der Waals superlattices.
  \emph{Nat. Commun.} \textbf{2018}, \emph{9}, 1413\relax
\mciteBstWouldAddEndPuncttrue
\mciteSetBstMidEndSepPunct{\mcitedefaultmidpunct}
{\mcitedefaultendpunct}{\mcitedefaultseppunct}\relax
\EndOfBibitem
\bibitem[Shinozaki \latin{et~al.}(2021)Shinozaki, Muto, Kitada, Nakajima,
  Delbecq, Yoneda, Takeda, Noiri, Ito, Ludwig, \latin{et~al.}
  others]{shinozaki2021gate}
Shinozaki,~M.; Muto,~Y.; Kitada,~T.; Nakajima,~T.; Delbecq,~M.~R.; Yoneda,~J.;
  Takeda,~K.; Noiri,~A.; Ito,~T.; Ludwig,~A., \latin{et~al.}  Gate voltage
  dependence of noise distribution in radio-frequency reflectometry in gallium
  arsenide quantum dots. \emph{Appl. Phys. Express} \textbf{2021}, \emph{14},
  035002\relax
\mciteBstWouldAddEndPuncttrue
\mciteSetBstMidEndSepPunct{\mcitedefaultmidpunct}
{\mcitedefaultendpunct}{\mcitedefaultseppunct}\relax
\EndOfBibitem
\end{mcitethebibliography}
\end{document}